# Comment on 'Intense picosecond x-ray pulses from laser plasmas by use of nanostructured "velvet" targets'

P. P. Rajeev and G. R. Kumar*
Tata Institute of Fundamental Research, Homi Bhabha Road, Mumbai 400 005, India.
(Dated: April 8, 2002)

Kulcsar et al.[1] recently reported 50-fold enhancement in x-ray yields from nickel 'velvet' plasmas produced by a high contrast, intense picosecond laser. They cited anisotropic dielectric constant, critical surface rippling and increased absorption length [2] as the causes of the enhancement. These were purely qualitative arguments and the remarkable enhancement was not quantitatively substantiated.

In this comment, we demonstrate that the enhanced fields due to surface structures can explain the observed enhancements. Modification of electric field by surface protrusions has been well studied in recent past [3–5] in the low intensity regime. Further, the field enhancement due to a plasmon resonance is a major source of hot electrons in a cluster plasma [6].

The surface of a 'velvet' target is modelled as a collection of hemispheroids of permittivity $\epsilon$, embedded on a flat substrate kept in vacuum. A p-polarized wave of amplitude $E$ is incident at an angle $\theta$ to the major axis of the spheroid. We assume, for simplicity, that the field along the major axis alone contributes to the enhancement. The resultant electric field in the vicinity of the hemispheroid becomes $\mathbf{E_r} = \mathbf{E_L^{out}} + E\cos\theta \hat{x}$[4], where $\mathbf{E_L^{out}}$ is the locally enhanced field and $E\cos\theta$ is the horizontal component of the incident electric field. For a long spheroid, as here, the enhancement is prominent towards the tip of the structure and is given by $\mathbf{E_L^{out}} = [L_\parallel^{out}\hat{\eta} + L_\perp^{out}\hat{\xi}]E\sin\theta$, where $L_\parallel^{out}$ and $L_\perp^{out}$ are the local field correction factors ($L_\parallel^{out} = L_\perp^{out}/\epsilon$, and is absent for metals). The effective intensity at the tip is

$$I_r = I_{in}[(L_\perp^{out})^2 \sin^2\theta + \cos^2\theta] \qquad (1)$$

The local field correction factors depend critically on $\epsilon$ and $\frac{a}{b}$, the aspect ratio of the spheroid, as illustrated in Fig. 1(inset). $L_\perp^{out} = 16$ for nickel at $\frac{a}{b} = \frac{800}{35}$ and $\theta = 20°$, at $\lambda = 1\mu$ (as in [1]), giving $I_r \sim 30 I_{in}$. This enhanced intensity causes extra absorption in a plasma sheath covering the structure. Fig. 1 provides a comparison of the original data in Fig. 2(b) of [1] with the data obtained by rescaling the points for 'velvet' target by $I_r$, and they are in good agreement as indicated by the extended linear fit for the flat target. Thus, the velvet target is shown to be equivalent to a flat target with rescaled intensity.

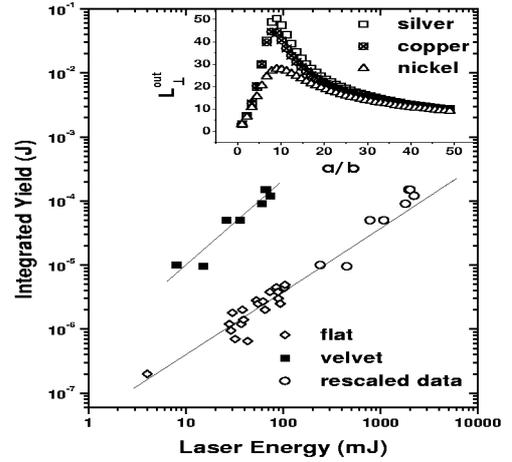

FIG. 1: Comparison of the rescaled data with the original by Kulcsar et al. Inset : $L_\perp^{out}$ vs. $\frac{a}{b}$ for different metals

This accordance is remarkable, as the model is an extremely simplified one. Ideally Helmholtz equation should be used since $\lambda \sim a$. The field interactions among the spheroids within the focal spot complicates the problem further. The electric field resonance within the plasma spheroid should be considered for long pulse durations. The key problem in the intense field regime is due to a drastically changing $\epsilon$ during the pulse as it severely affects enhancement factors for low $b$ values. Lastly, a variety of nonlinear effects which occur in the intense field regime get enhanced by different magnitudes and they ought to be summed carefully. This simple model is, however, attractive as it suggests that the field modifications near surface structures could cause excess absorption and provides a good pointer for designing better targets (with $a/b = 10, b = 10nm$ a silver 'velvet' target experiences 16 times larger $I_r$ than a nickel 'velvet' target).

PPR acknowledges discussions with V. Kumarappan *electronic mail: grk@tifr.res.in

[1] G. Kulcsar *et al.*, Phys. Rev. Lett. **84**, 5149(2000)
[2] M.M. Murnane *et al.*, Appl. Phys.Lett. **62**, 1068(1993)
[3] J. Gersten, and A. Nitzan, J. Chem. Phys. **73**, 3023(1980)
[4] G. T. Boyd *et al.*, Phys. Rev. B **30**, 519(1984)
[5] M.B Mohamed *et al.*, Chem. Phys. Lett., **317**, 517(2000)
[6] T. Ditmire *et al.*, Phys. Rev. A **57**, 369(1998)